\documentstyle[12pt,epsf]{article}

\begin{document}

\begin{titlepage}

\null
\begin{flushright}
CERN-TH.7149/94\\
ILL-(TH)-94-26
\end{flushright}
\vspace{20mm}

\begin{center}
\bf\Large
Phase Structure of Four Dimensional
Simplicial Quantum Gravity
\end{center}

\vspace{5mm}

\begin{center}
{\bf S. Catterall\footnote{Permanent address: Physics Department, Syracuse
University, Syracuse, NY 13244}}\\
TH-Division, CERN CH-1211,\\
Geneva 23, Switzerland.\\
{\bf J. Kogut}\\
Loomis Laboratory, University of Illinois at Urbana,\\
1110 W. Green St, Urbana, IL 61801.\\
{\bf R. Renken}\\
Department of Physics, University of Central Florida,\\
Orlando, FL 32816.
\end{center}

\begin{center}
\today
\end{center}

\vspace{10mm}
      
\begin{abstract}
We present the results of a high statistics Monte Carlo study of a 
model for four
dimensional euclidean quantum gravity based on summing over
triangulations. We show evidence for two phases; in one there is
a logarithmic scaling of the mean linear extent with
volume, whilst the other exhibits power law behaviour with exponent $\frac 1 2$.
We are able to extract a finite size scaling exponent
governing the growth of the susceptibility peak.
\end{abstract}

\vfill
\begin{flushleft}
CERN-TH.7149/94\\
January 1994
\end{flushleft}

\end{titlepage}

\section*{Introduction}

In this article we report briefly on some new results for a model whose
partition function is constructed from a sum over all
random triangulations of the four dimensional sphere. Such a model
has been recently studied by other groups \cite{mig,amb,varsted,brug} and is
a candidate for a regularised quantum theory of euclidean gravity 
(for fixed euler number).
Specifically, the model we have looked at is defined from the
partition function.
\begin{equation}
Z=\sum_{T, \chi =2} e^{-\kappa_4 N_4 + \kappa_0 N_0}
\end{equation}
The first term in the action $N_4$ is just the number of four simplices
in the triangulation $T$ and this allows us to identify the corresponding
coupling $\kappa_4$ as a bare cosmological constant. The second term plays the
role of the integrated Ricci scalar ($N_0$ is just the number of vertex labels
or points in the manifold). This correspondence is clear (classically) from
the usual Regge formula for the curvature in terms of the deficit angle
with the extra constraint that the four simplices are all
considered equilateral
\begin{equation}
\int d^4 z \sqrt{g} R = 4\pi\left(N_0+N_4-2\right)-10\cos^{-1}{\left(1\over 4
\right)}
\end{equation}
Notice, also, that the formula for the local curvature at a hinge (triangle
$ijk$) takes the form
\begin{equation}
r_{ijk}=2\pi - \cos^{-1}{\left(1\over 4\right)} n_4^{ijk}
\end{equation}
So, if the volume is bounded the number of four simplices sharing the triangle
$n_4^{ijk}$ is necessarily also bounded. This automatically ensures that
the model is well defined at finite volume -- it is a dynamical
question as to whether the unboundedness problems return on taking the
large volume limit.

The analogous model in two dimensions has been studied extensively, both
in the continuum using analytic methods and via the triangulated
lattice prescription, see for example the review \cite{david}. It seems clear
that at least for central charges less than unity, the sum over triangulated
graphs correctly mimics the continuum functional integrals including the
conformal anomaly. In four dimensions it is not at all clear that a simple
generalisation, such as the one described above, is sufficient to explore
the space of metrics. However, at this stage, it provides a convenient
ansatz which may be studied using numerical simulation.

In practice, we choose to work at quasi fixed volume -- that is we tune the
cosmological constant $\kappa_4$ to fix the mean volume $\overline N_4$ to
some target value $V$. To remove some of the fine tuning problems
associated with this procedure we add a 
small correction term to the action \cite{mig}
of the form
\begin{equation}
\delta S=\gamma\left(N_4-V\right)^2
\end{equation}
We have verified that expectation values computed from this modified action
do not depend on the coupling $\gamma$, although their errors
increase as $\gamma\to 0$. In practice we have set $\gamma =0.005$ which
yields an average fluctuation in the volume in the region of a fraction
of a percent.

At fixed volume, the model depends on just one parameter -- the node coupling
$\kappa_0$. As we have argued this is just inversely related to the bare
Newton constant. Questions of renormalisability of continuum gravity are
then replaced by the search for continuous phase transitions in this one
dimensional parameter space. 
The hope would be that in the vicinity of
such a point, it might be possible to recover
a nonperturbative continuum limit for quantum gravity.

We have used a Monte Carlo algorithm to sample the triangulation space
of the model. This procedure is based on a set of `moves' or local
retriangulations of the manifold, which are ergodic in the grand canonical
ensemble used here. This means that it is possible
to reach any triangulation from any other by sequences of these
moves. In $d$ dimensions there are $d+1$ moves \cite{mig,amb} which may
be pictured as the trial substitution of an $i$-subsimplex by its
dual $(d-i)$-subsimplex. Our code is written in such a way as to make the
dependence on dimension $d$ trivial -- it enters only as an input parameter
to the program \cite{us}. This allows us to test the code in two dimensions.
Finally, it is very important to ensure the update satisfies a detailed
balance condition in order that the contribution of a given
triangulation depends only on its Boltzmann weight. 

We have simulated systems with sizes ranging from $V=500$ to $V=32000$, although
the majority of our high precision data is confined to lattices of $V=8000$
and smaller. On the latter we have accumulated up to $8\times 10^6$
sweeps at each value of $\kappa_0$ ($1$ sweep corresponds to $V$ attempted
elementary moves). 

\section*{Results} 

\begin{figure}
\begin{center}
\leavevmode
\epsfxsize=400pt
\epsfbox{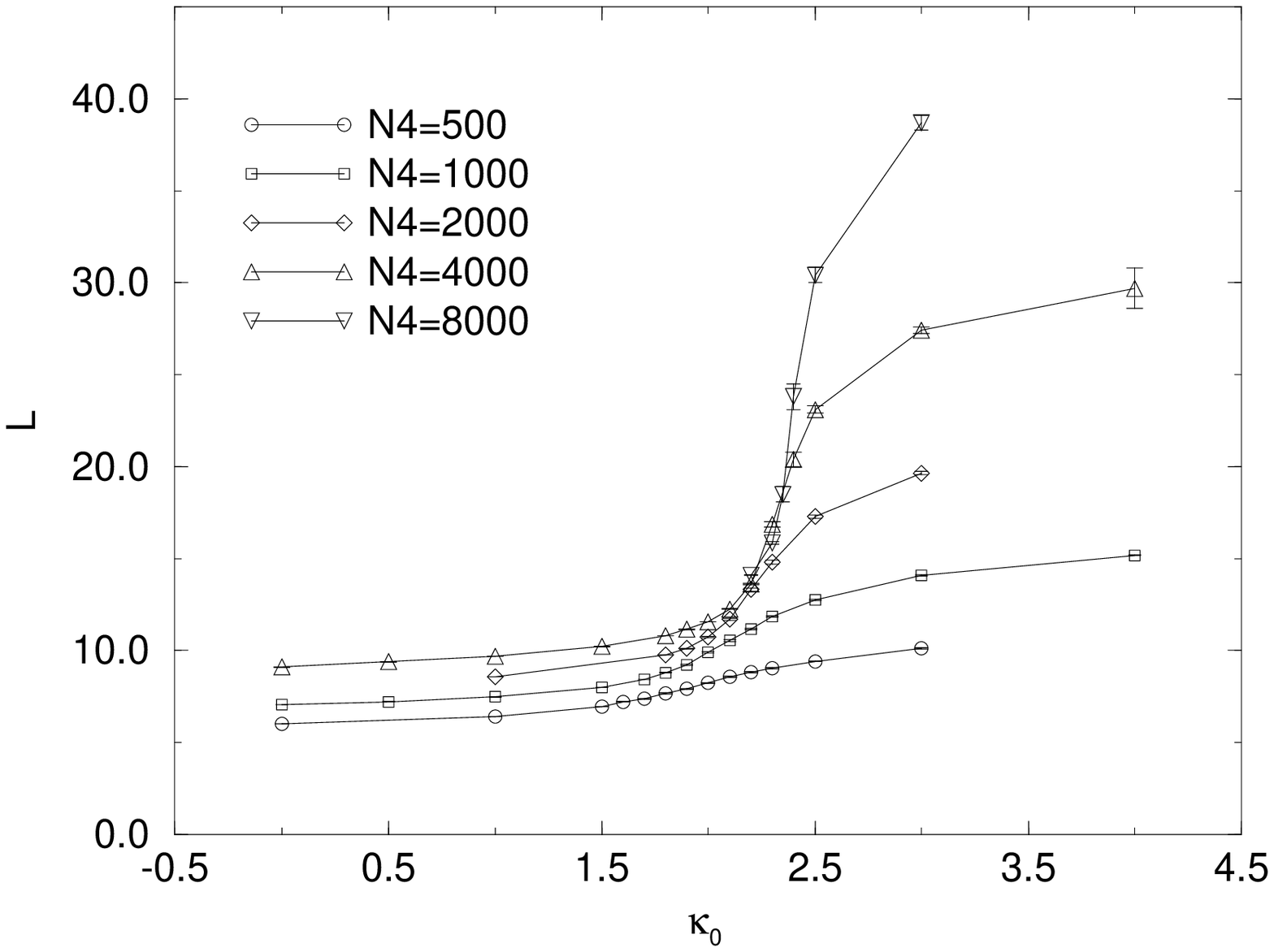}
\caption{Mean extent}
\label{fig1}
\end{center}
\end{figure}

A crude order parameter which may be used to distinguish 
between different phases
of the model is the average {\it intrinsic} linear extent of the system.
This may be defined as follows
\begin{equation}
L=\left\langle {1\over V^2}\sum_{ij}^V d\left(i,j,T\right)\right\rangle_T
\end{equation}
The distance $d(i,j,T)$ on a given triangulation $T$ is just the minimal
number of steps on the dual lattice between the simplex $i$ and another
simplex $j$. Fig.\ \ref{fig1} shows a plot of this for lattice volumes ranging
from $V=500$ to $V=8000$. Clearly, at small $\kappa_0$ the typical
manifolds are very collapsed and $L$ increases only very slowly with
mean volume. Indeed, for fixed $\kappa_0$ in this range, it is possible to
fit the data quite successfully ($\chi^2\sim O(1)$) to a 
logarithm of the volume $V$.

\begin{equation}
L=a\left(\kappa_0\right)+b\left(\kappa_0\right)\ln V
\end{equation}
 
If we care to define a fractal dimension $d_F$
for the system from the formula $l\sim N^{1\over d_F}$, this would
imply $d_F=\infty$. Power fits give order of magnitude worse
values for the goodness-of-fit parameter $\chi^2$.

Conversely, at large $\kappa_0$ the manifolds are extended with $L$
scaling like a power of $V$ (at least for volumes up to $V=8000$). 
Logarithmic fits perform much more poorly here. In fact at $\kappa_0=
3.0$ a fit yields $d_F=2.08(4)$ and at $\kappa_0=4.0$ a value of $d_F=1.98(2)$.
Both fits have $\chi^2$ of order the number of degrees of freedom. This
fractal dimension seems to indicate that typical
configurations in this phase resemble some sort of branched polymer.

We have also measured the fluctuations (per unit volume) in the
mean extent $L$. 

\begin{equation}
\chi\left(L\right)={1\over V}\left(\left\langle L^2\right\rangle-
\left\langle L\right\rangle^2\right)
\end{equation}

\begin{figure}
\begin{center}
\leavevmode
\epsfxsize=400pt
\epsfbox{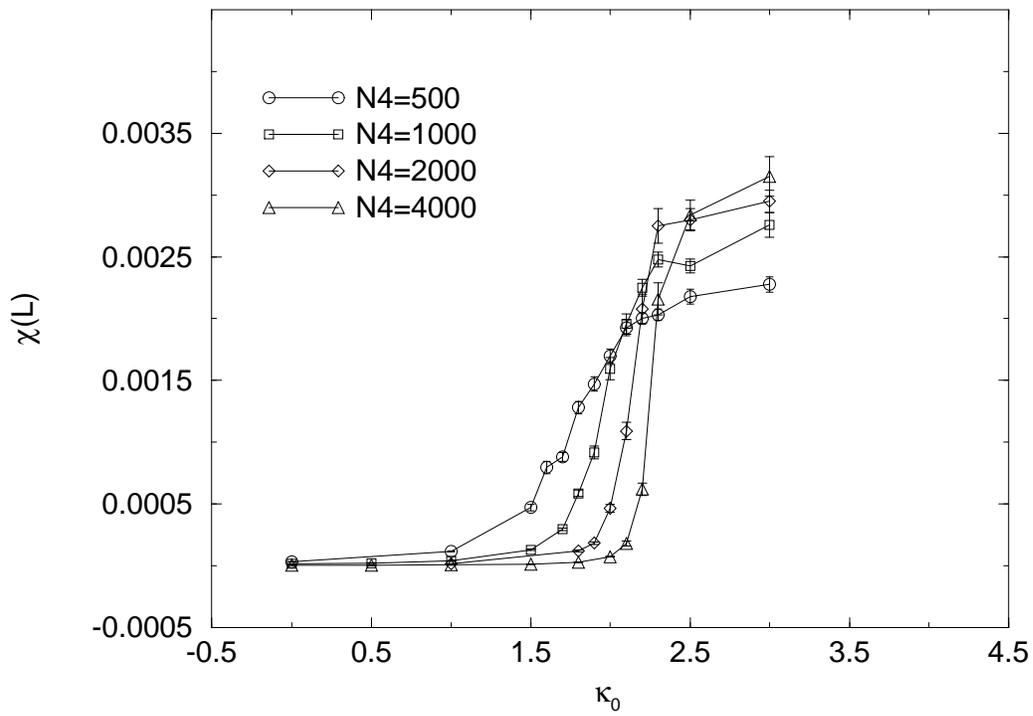}
\caption{Fluctuation in mean extent}
\label{fig2}
\end{center}
\end{figure}

For small $\kappa_0$ this quantity is small and
decreases with increasing volume. Conversely, at large $\kappa_0$ it
plateaus at a value several orders of magnitude larger. Furthermore
this asymptotic value appears to {\it increase} with volume $V$.
It thus functions perhaps more effectively as an order parameter, being
(for large volume) zero for a range of small couplings and
very large (perhaps infinite as $V\to\infty$) for all large couplings.
The transition between these two regions appears to become
ever more sharp with increasing volume as fig. \ \ref{fig2} indicates.
It is tempting to identify the crossover point as a (pseudo) critical
coupling $\kappa_0^c\left(V\right)$.

A more conventional indicator of phase transitions is the specific heat,
which in this case is just the fluctuation in the number of nodes
\begin{equation}
\chi\left(N_0\right)={1\over V}\left(\left\langle N_0^2\right\rangle -
\left\langle N_0\right\rangle^2\right)
\end{equation}

\begin{figure}
\begin{center}
\leavevmode
\epsfxsize=400pt
\epsfbox{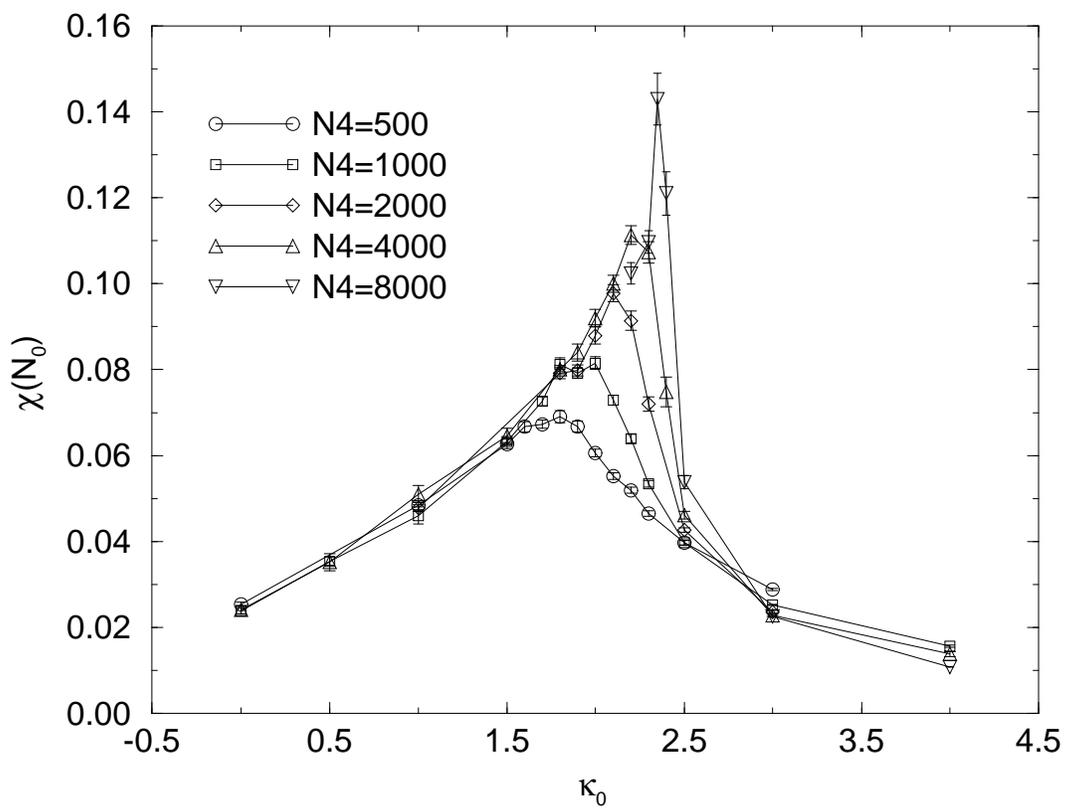}
\caption{Node susceptibility}
\label{fig3}
\end{center}
\end{figure}

This quantity is proportional to the integrated 2-pt function
for the lattice scalar curvature 
and hence an enhancement would
signal the onset of long range curvature correlations. 
In fig.\ \ref{fig3} we see 
a sharp peak developing in the vicinity of $\kappa_0\sim 2$, which grows,
shifts and narrows with increasing volume. Furthermore,
the position of this peak coincides with the jump in
the geodesic susceptibility $\chi\left(L\right)$. There is a
corresponding increase in the autocorrelation time close to
the peak, which we estimate grows roughly linearly with the volume.
This then is our best evidence for a continuous phase transition in the system.

\begin{figure}
\begin{center}
\leavevmode
\epsfxsize=400pt
\epsfbox{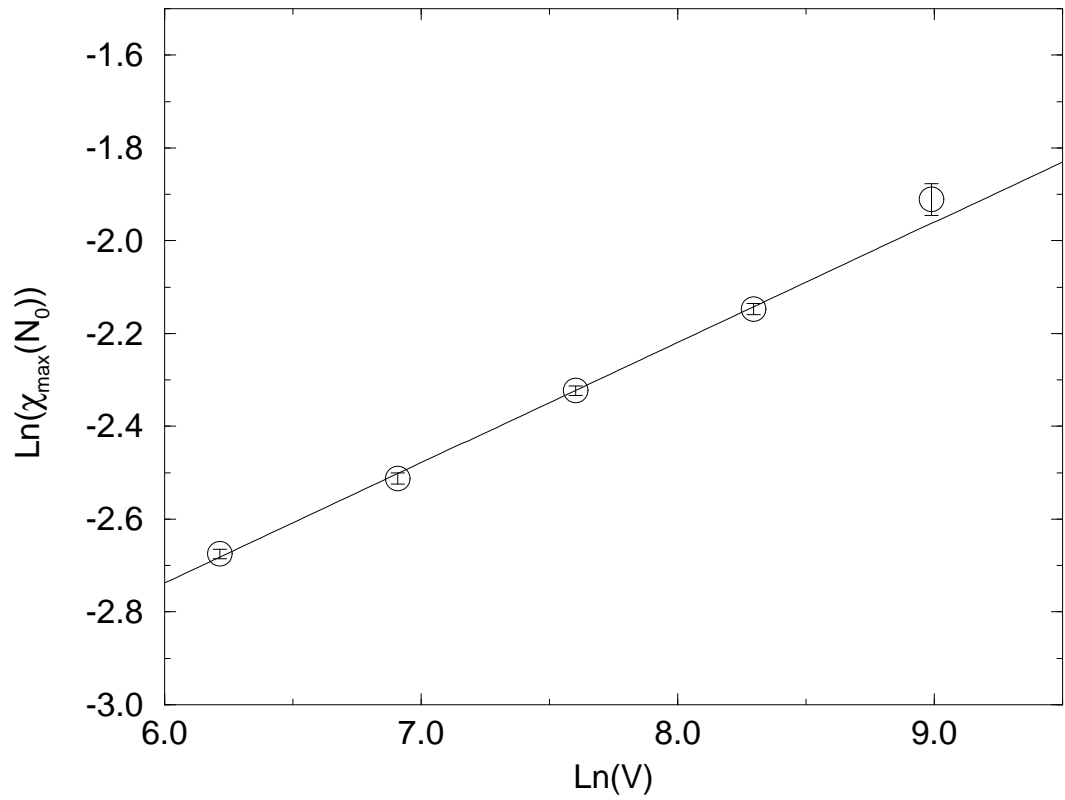}
\caption{Log max node susceptibility}
\label{fig4}
\end{center}
\end{figure}

In systems coupled to two dimensional gravity
the expectation value of some integrated quantity
scales as a power of the volume, the exponent being related to the
(gravitationally dressed) anomalous dimension of the matter field. 
Motivated by this we have looked for a power law scaling of
the node susceptibility peak. Indeed, the data shown in
fig.\ \ref{fig4} appears quite
consistent with such a scaling scenario with a power $\Delta=0.259(7)$
($\chi^2\sim 1.2$, 29\%). 
If, in addition,
we assume conventional finite size scaling we might
tentatively associate this power with
$\Delta={\alpha\over {\nu d_F}}$ ($\alpha$ and $\nu$ would then be
the susceptibility and correlation length exponents). However the
assumptions needed to establish this result may not be applicable for
these systems.

Fig. \ \ref{fig3} makes it clear that
the peak in the node susceptibility is moving rather
rapidly with increasing volume which opens up the (rather unpleasant!)
scenario that it may diverge in the infinite volume limit. Indeed,
on cursory inspection it appears possible that the pseudo critical
coupling might have a component proportional to the logarithm
of the mean volume $V$. The putative phase
transition would then not survive the infinite volume
limit and for any $\kappa_0$ the large volume
behaviour of the model would correspond to the compact, degenerate phase.
On the other hand the slow convergence could simply be the
result of a small shift exponent, which might not be unreasonable
given the very large intrinsic dimensionality of the compact phase.
 
The raw data is simply not good enough to resolve this crucial
issue and so we have 
used multihistograming techniques \cite{swend}
to interpolate the susceptibility in the vicinity
of the peak.
The errors in the pseudo critical couplings $\kappa_0^c(V)$
were then assessed by performing 
several such reconstructions using simulations at different
neighboring couplings $\kappa_0$.

\begin{figure}
\begin{center}
\leavevmode
\epsfxsize=400pt
\epsfbox{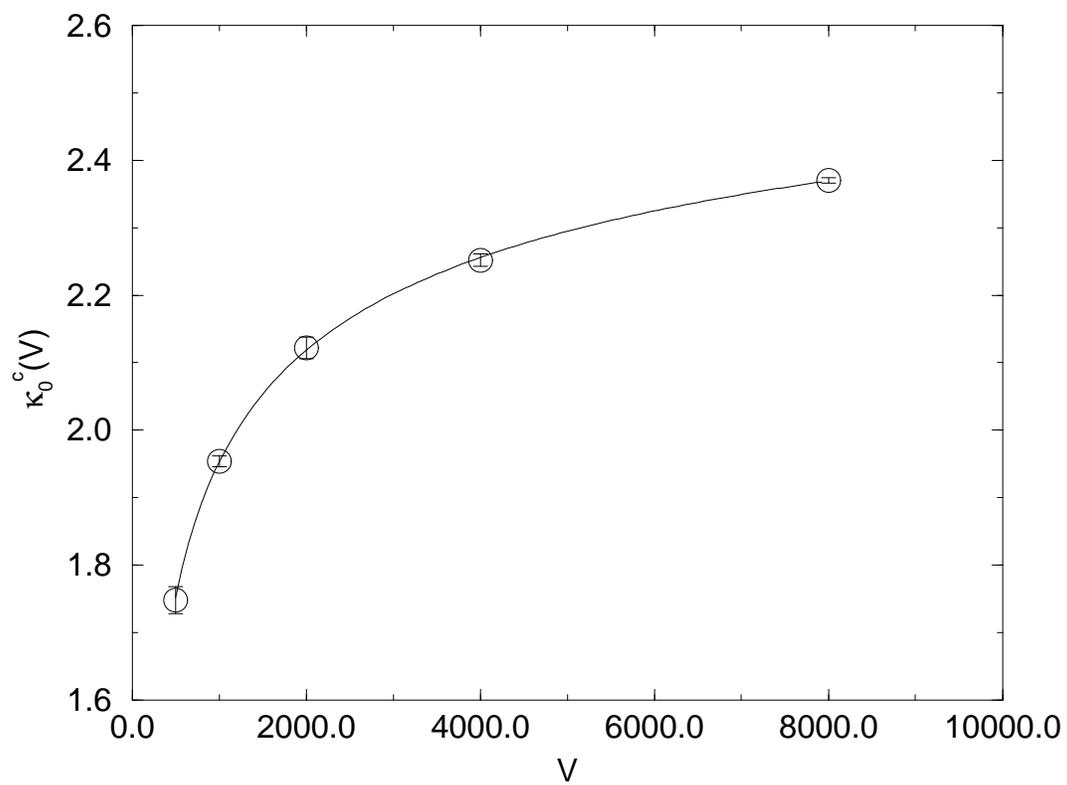}
\caption{Critical coupling}
\label{fig5}
\end{center}
\end{figure}

To try to distinguish between these
two scenarios we have attempted to fit the pseudo critical
coupling $\kappa_0^c(V)$ data with both a power law form 
\begin{equation}
\kappa_0^c\left(V\right)=a+bV^\omega
\end{equation}
and a modified logarithm
\begin{equation}
\kappa_0^c\left(V\right)=a+b\left(\ln V\right)^\omega
\end{equation}
In both cases, the {\it converging} fit ($\omega < 0$) was better
and yielded statistically compatible estimates for the infinite
volume critical coupling $\kappa_0^c(\infty )$.
Fig.\ \ref{fig5} shows the data, together with the result of a power fit
with $a=2.9(2)$, $b=-6.3(16)$ and exponent $\omega=-0.27(6)$. The fit
yields a $\chi^2$ per degree of freedom of $0.2$ at 85\% confidence.
Conventional finite size scaling would then identify
the exponent $\omega=-{1\over \nu d_F}$.
By contrast the simple {\it diverging} log fit ($\omega=1$)
gave $\chi^2\sim O(10)$. However, one must be careful in drawing
too many conclusions from this. If one drops the two smallest
lattices an equally good fit can be got from the simple logarithm.

\section*{Outlook}

We have presented some new results for a four dimensional model of
dynamical triangulations. Emphasis has been placed on gaining
some quantitative understanding of the thermodynamic properties
near the putative phase transition. This has
entailed accumulating two orders of magnitude
more Monte Carlo data than typical earlier studies. We are able, for the
first time, to convincingly resolve a finite size scaling
for the specific heat peak. This exhibits a power law growth and
we have been able to estimate the exponent rather accurately
$\Delta=0.259(7)$. The
value is inconsistent with that appropriate to a first order
transition ($\Delta=1.0$). As a further check,
we have histogramed the action and have seen no evidence for
the two peak structure typical of first order transitions.
This evidence then favours a continuous transition.

However, as we have emphasised, the position of this specific heat peak 
moves rather rapidly with volume, leaving open the possibility that the
pseudo critical coupling diverges with volume, the phase
transition disappearing in the infinite volume limit. We have
attempted  to distinguish between a slow power law convergence 
of the critical coupling (conventional scenario) from a 
logarithmic or power law divergence. Since we have high statistics
data with good resolution of the peak we found that histogram
reconstruction was a reliable tool for the critical region
and allowed us to follow the evolution of the pseudo critical
couplings in more detail.
The tentative conclusion from this analysis is that our data favour
the conventional scenario with a finite value for the
infinite volume critical
coupling $\kappa_0^c\left(\infty\right)=2.9(2)$. 

This 
transition then separates two very different phases; at small  
coupling $\kappa_0$ the entropy of small, highly connected manifolds
with infinite fractal dimension dominates, at large $\kappa_0$, the
action favours extended structures with dimension two. In this
latter phase the fluctuations in the geodesic extent are always
large and indeed yield large autocorrelation times for
all $L$-dependent observables. This phase can then
presumably be identified as consisting of randomly branching polymers.  

It is clearly important to strengthen these conclusions by extending
the analysis to larger systems. This is
particularly true for the determination of the
shift exponent (and the question of the finiteness of
$\kappa_0^c(\infty )$). The present maximum volume
was determined by the large autocorrelation time $O(10^4)$ sweeps
for $V=8000$
encountered near the
transition. We would hope to address this in the near
future. It will then be necessary to begin to understand the
relevance of the critical model to a continuum theory of gravity. 

This work was supported, in part, by NSF grant PHY 92-00148. Some
calculations were performed on the Florida State University Cray YMP.

\vfill
\newpage

\end{document}